\documentclass[12pt]{article}
 \pdfoutput=1

\usepackage{amsmath,amssymb,amsfonts}
\usepackage{graphicx}
\usepackage{color}
\definecolor{darkblue}{rgb}{0.1,0.1,.7}
\usepackage[colorlinks, linkcolor=darkblue, citecolor=darkblue, urlcolor=darkblue, linktocpage]{hyperref} 
\usepackage[square, comma, compress,numbers]{natbib}

\usepackage{mathrsfs}
\usepackage[margin=10pt,font=small,labelfont=bf]{caption}
\newcommand{\hhref}[1]{\href{http://arxiv.org/abs/#1}{arXiv:#1}}
\newcommand{\del}{\partial}

\newcommand{\eref}[1]{Eq.~(\ref{e.#1})}

\newcommand{\cref}[1]{Chapter~\ref{c.#1}}

\newcommand{\barray}{\begin{eqnarray}}
\newcommand{\earray}{\end{eqnarray}}
\newcommand{\nn}{\nonumber \\}

\newcommand{\beq}{\begin{equation}} 
\newcommand{\eeq}{\end{equation}} 
\newcommand{\ba}{\begin{array}}  
\newcommand{\ea}{\end{array}} 
\newcommand{\bea}{\begin{eqnarray}}  
\newcommand{\eea}{\end{eqnarray} }  
\newcommand{\be}{\begin{eqnarray}}  
\newcommand{\ee}{\end{eqnarray} }  
\newcommand{\bal}{\begin{align}}
\newcommand{\eal}{\end{align}}   
\newcommand{\bi}{\begin{itemize}}  
\newcommand{\ei}{\end{itemize}}  
\newcommand{\ben}{\begin{enumerate}}  
\newcommand{\een}{\end{enumerate}}  
\newcommand{\bc}{\begin{center}}
\newcommand{\ec}{\end{center}} 
\newcommand{\bt}{\begin{table}}
\newcommand{\et}{\end{table}}  
\newcommand{\btb}{\begin{tabular}}
\newcommand{\etb}{\end{tabular}}  
\newcommand{\bvec}{\left ( \ba{c}}
\newcommand{\evec}{\ea \right )}

\def\cl{{\mathcal L}}

\def\ra{\rangle}
\def\la{\langle}  

\def\pa{\partial}  

\newcommand{\im}{{\mathrm{Im}} \,}
\newcommand{\tr}{\mathrm T \mathrm r}

\def\hc{{\rm h.c.}} 

\newcommand{\eps}{\epsilon}

\newcommand{\qstate}[1]{| #1 \rangle}
\newcommand{\ostate}[1] {\langle #1 |}
\def\calA{{\mathcal A}}

\begin{document}
\begin{titlepage}
\vspace{-1cm}
\begin{flushright}
\small
LPTENS--12/05 \\
LPT-ORSAY 12-13
\end{flushright}
\vspace{0.2cm}
\begin{center}
{\Large \bf What if the Higgs couplings to W and Z bosons \\ are larger than in the Standard Model? }
\vspace*{0.2cm}
\end{center}
\vskip0.2cm

\begin{center}
{\bf  Adam Falkowski$^{a}$, Slava Rychkov$^{b,c}$, Alfredo Urbano$^{b}$}

\end{center}
\vskip 8pt

\begin{center}
{\it $^{a}$ Laboratoire de Physique Th\'eorique d'Orsay, UMR8627--CNRS,\\ Universit\'e Paris--Sud, Orsay, France}\\
{\it $^{b}$  Laboratoire de Physique Th\'eorique, \'Ecole Normale Sup\'erieure, Paris, France} \\  
{\it $^{c}$ Facult\'e de Physique, Universit\'e Pierre et Marie Curie, Paris, France} 
\end{center}


\vspace*{0.3cm}

\vglue 0.3truecm

\begin{abstract}
\vskip 3pt \noindent 
We derive a general sum rule relating the Higgs coupling to W and Z bosons to the total cross section of longitudinal gauge boson scattering in  $I=0,1,2$ isospin channels. 
The Higgs coupling larger than in the Standard Model implies enhancement of the $I=2$ cross section.    
Such an enhancement could arise if the Higgs sector is extended by an isospin-2 scalar multiplet  including a doubly charged, singly charged, and another neutral Higgs.  

\end{abstract}
\vspace{.2in}
\vspace{.3in}
\hspace{0.7cm} February 2011

\end{titlepage}

\newpage

{\bf Introduction.} 
The LHC is on its to way to discovering a Higgs boson and measuring or constraining its couplings to other Standard Model (SM) particles. 
The coupling to W and Z bosons are particularly important because they control the high-energy behavior of the scattering amplitude of longitudinally polarized electroweak gauge bosons. 
Theoretical constraints on that coupling were previously discussed in Ref.~\cite{Low:2009di} in the framework of Strongly Interacting Light Higgs (SILH)  \cite{Giudice:2007fh}, where an approximately elementary Higgs doublet arises as a pseudo-Goldstone boson in a strongly interacting sector. 
In this letter we revisit this question, without making reference to an elementary Higgs doublet field. By the Higgs boson we simply mean any light neutral scalar particle with custodial isospin 0 and a significant coupling to W and Z.
 
Parametrizing the Higgs coupling to W and Z as\footnote{The fact that only one parameter $a$ controls the Higgs coupling to both W and Z boson is the consequence of assuming custodial symmetry, which is strongly suggested by electroweak precision tests. } 
\beq
\label{e.hvv}
\cl_{hVV} = a \, {h \over v}  \left ( 2 m_W^2W_\mu^+ W_\mu^- + m_Z^2 Z_\mu Z_\mu \right ) 
\eeq 
we will see that there is a sum rule relating the coefficient $a$ to a linear combination of the total cross sections in different isospin channels of longitudinal  electroweak gauge boson scattering: 
\beq
\label{e.master}
1 - a^2  = {v^2 \over 6 \pi}  \int_0^\infty {ds \over s}  
\left ( 2 \sigma_{I=0}^{\rm tot}(s)+  3  \sigma_{I=1}^{\rm tot}(s)  -  5 \sigma_{I=2}^{\rm tot}(s) \right ).
\eeq 
The equality holds for a light Higgs boson and in the limit of vanishing electroweak gauge couplings, $g,g' \to 0$. 
The SM predicts $a =1$.
Intriguingly, CMS recently reported an excess of Higgs-like events in the diphoton channel  produced in association with  2 forward jets \cite{cmsgaga}. 
This may possibly be interpreted as an enhancement of the vector-boson-fusion Higgs production mode and therefore a hint for $a>1$.
From \eref{master} it is clear that the Higgs coupling to W and Z {\em exceeding} the SM value implies that the cross section in the isospin-2 channel dominates over the remaining 2 channels, at least for a certain range of invariant mass $s$.   
The simplest way to satisfy the sum rule with $a > 1$  is by introducing a resonance in the isospin-2 channel. 

{\bf Derivation of the sum rule.} 
Below we derive \eref{master}. We will use the equivalence theorem, where the scattering amplitudes of $W^+_L,W^-_L,Z_L$ are approximated by scattering amplitudes of a triplet massless ``pions" $\pi^a$ parametrizing the coset space of the $SU(2) \times SU(2)/SU(2)_V$ non-linear sigma model. 
The $SU(2)_L \times U(1)_Y$ subgroup of $SU(2) \times SU(2)$ is weakly gauged by the electroweak vector fields. 
At low energies, where the only degrees of freedom are those of the SM (including the Higgs boson $h$), the most general Lagrangian with these symmetries can be parametrized as
\beq
\label{e.LEL}
\cl = {1 \over 2} (\pa_\mu h)^2 - V(h)  
+ {v^2 \over 4} \tr \left [D_\mu U^\dagger D_\mu U  \right ] \left (1 + 2 a {h \over v} + b {h^2 \over v^2} + \dots \right ), 
\qquad U = e^{i \pi^a \sigma^a/v}. 
\eeq 
where $D_\mu U = \pa_\mu U - i (g/2) \sigma^a L_\mu^a U  + i (g'/2) B_\mu U \sigma^3$, but we will work in the limit $g,g' \to 0$. Using isospin and crossing symmetry, the basic pion scattering amplitude must have the form 
\beq
T_{ab\to cd}=\ostate{cd} S \qstate{ab}
= A_s \delta^{a b} \delta^{c d} + A_t \delta^{a c} \delta^{b d} +  A_u \delta^{a d} \delta^{b c} ,  
\label{e.basic}
\eeq 
where $A_s = A(s,t,u)$, $A_t = A(t,s,u)$, $A_u = A(u,t,s)$, and $s+t+u=0$ for massless pions. At low energies, the amplitudes can be computed from the Lagrangian \eref{LEL}, which gives:
\beq 
\label{e.lowAmp}  
A(s,t,u) = {s \over v^2}\left(1 - a^2 {s \over s - m_h^2}\right)  \approx  \left (1 - a^2 \right) {s \over v^2} . 
\eeq  

In what follows we will use the amplitudes projected on definite isospin of the initial and final 2-pion state (see Appendix for a recap of isospin decomposition). In terms of the function $A(s,t,u)$ these projections are given by:  
\beq 
\label{e.isoAmp}  
T_0(s,t) = 3 A_s + A_t + A_u\,, 
\quad   
T_1 (s,t) =  A_t - A_u\,,
\quad   
T_2 (s,t) =   A_t + A_u\,. 
\eeq 
The idea of the proof is to consider a certain contour integral of a general linear combination of forward isospin amplitudes in the complex plane $s$
\beq
I_{\calA} = {1 \over 2 \pi i} \int_{C_0} ds \,   {\calA(s) \over s^2}  \qquad \calA(s) = \sum_I w_I T_I (s)\equiv \sum_I w_IT_I(s,t=0)  
\eeq   
where $C_0$ is a small circle around $s = 0$. Here $T_I (s,t)$ are amplitudes in the full UV-complete theory. Since at low energies they can be approximated using \eref{lowAmp}, straightforward integration yields
\beq
\label{e.ia1}
I_\calA = (1-a^2) \left (2 w_0 + w_1 - w_2 \right)/v^2.
\eeq 
On the other hand, deforming the circle to a big one and around the cuts on the real axis, as is customary in this kind of arguments \cite{Adams:2006sv}, we get
\beq
I_\calA = {1 \over 2 \pi i} \left (  \int_{C_\infty} {ds \over s^2}  \calA(s)  + \int_0^\infty {ds \over s^2} [\calA(s+i \eps) -\calA(s- i \eps)] -  \int_0^{\infty} {ds \over s^2}  [\calA(- s- i \eps) - \calA(-s+i \eps)]  \right )
\eeq
where  $C_\infty$ is the circle at infinity. We are thus relating IR physics parameter $a$ to the unknown (but potentially measurable) amplitudes at high energies, which far away from $s= 0$ can no longer be approximated  using  \eref{lowAmp}.

Next, we would like to express the contribution of the negative $s$ cut in terms of amplitudes evaluated at positive $s$. 
Crossing symmetry implies 
\beq
T_{ab\to cd}(-s,t=0)=T_{ad\to cb}(s,t=0). 
\eeq
 This means that $A_{s}$ goes to $A_u$ under $s\to-s$. Using \eref{isoAmp} and its inverse, it is easy to get the corresponding relation for $T_I$'s:
\begin{align}
T_0(-s)&\textstyle=\frac {1}3 T_0(s)-T_1(s)+\frac53 T_2(s) ,  \nonumber\\
T_1(-s)&\textstyle=-\frac {1}3 T_0(s)+\frac12 T_1(s)+\frac56 T_2(s) , \nonumber\\
T_2(-s)&\textstyle=\frac {1}3 T_0(s)+\frac12 T_1(s)+\frac16 T_2(s). 
\end{align} 
It is now possible to combine the two cuts into a single positive $s$ integral. We get:
\beq
I_\calA = {1 \over 2 \pi i }   \int_{C_\infty} {ds \over s^2}  \calA(s)  +  {2 w_0 + w_1 - w_2 \over \pi}\int_0^\infty {ds \over s^2}  \left [ {1 \over 3} \im T_0(s) + {1 \over 2} \im T_1(s) -  {5 \over 6 }\im T_2(s)\right ],
\label{e.ia2}
\eeq
Note that the weights $w_I$ factor out, and it's exactly the same factor that occurs in \eref{ia1}.  
The optical theorem relates the imaginary part of the forward amplitude to the total scattering cross section, $\im T_I(s) = s \sigma_I^{\rm tot}(s)$ for massless pions. 
Then, assuming that the integral over the big circle vanishes, that is $\calA(s)$ grows slower than $s$ at infinity,  and then  comparing \eref{ia1} and \eref{ia2}  we recover the sum rule \eref{master}.

A similar sum rule with $a = 0$ holds for pion-pion scattering in QCD, in which context it has been derived by Olsson \cite{Olsson}. It has also been given by Adler \cite{Adler}, in a reduced form as in Eq.~(\ref{eq:reduced}) below. See also \cite{Nieves:2011gb} for a recent discussion.  

{\bf Discussion.} 
Let us now comment on how general our result is. First of all, we are assuming that the electroweak symmetry breaking sector has a UV completion of a kind which makes it possible to speak of an analytic and unitary S-matrix even at energies much higher than the electroweak scale. This is a reasonably general assumption which should be valid for any field or string theory-type UV completion. 

Second, in the process of our proof we had to assume that $\calA(s)$ grows strictly slower than $s$ at infinity, to make the integral over the big circle vanish. This assumption requires caution, since the Froissart bound
does allow total cross section $O(\log^2s)$, which would correspond to a forward scattering amplitude $O(s\log^2 s)$. However, not all amplitudes are expected to grow maximally fast. 

Suppose first that the UV completion is strongly coupled. Then we can hope to apply the Regge theory, which predicts the high energy behavior of the amplitude at $s\gg t$ of the form
\beq
T_{AB\to CD}(s,t)\sim Z \gamma_{AC}\gamma_{BD}\, s^{\alpha(0)+\alpha't}\quad\text{(modulo factors of $\log s$)}
\eeq
where $\alpha(0)$ and $\alpha'$ are the intercept and the slope of the leading Regge trajectory which can be exchanged in the t-channel, $\gamma_{AB}$ and $\gamma_{CD}$ are the corresponding couplings, and $Z$ is a trajectory-dependent complex number \cite{Pomeron}. The amplitudes saturating the Froissart bound are those which can couple to the pomeron: the isospin 0 trajectory with quantum numbers of the vacuum and of intercept $\alpha(0)=1$. On the other hand, processes with nonzero isospin exchange (like $\pi^-p\to\pi^0n$ in QCD) have cross sections falling off like a power of energy (the Pomeranchuk theorem). 

Specializing to the pion scattering amplitude, the isospin 0 nature of the pomeron implies that $\gamma_{ac}=C\delta_{ac}$ and
\beq
T_{ab\to cd}(s,t=0) =  Z C^2\delta_{ac}\delta_{bd}\, s\, + \dots ,
\label{eq:Tregge}
\eeq
while the terms which exchange isospin 1 and 2 in the t-channel must be power suppressed. From here it is obvious that the fixed isospin amplitudes have identical leading $s$ behavior:
\beq
\label{eq:Tregge2}
T_I(s)\sim Z C^2 s,\qquad I=0,1,2 . 
\eeq
This causes problems in our proof, as both the contribution of the big circle is nonvanishing and the integrals over the negative and positive cuts are divergent before being combined (notice that generically $Im\, Z\ne0$). Fortunately, there is a simple way out. Let us use the freedom to choose the weights $w_I$ in our proof so that $w_0+w_1+w_2=0$ (but $2 w_0 + w_1 - w_2\ne0$). Then $\calA(s)$ grows slower than $s$ and our argument goes through. Thus our theorem holds also for strongly coupled UV completions, as long as the Regge theory applies. Incidentally, Eq.~(\ref{eq:Tregge2}) implies also the convergence of our sum rule (\ref{e.master}), as the factors multiplying the isospin cross sections sum up to zero.

For the weakly coupled UV completions the story complicates somewhat. Such UV completions can contain massive vectors of isospin 1 whose t-channel exchange gives the leading $s\gg M_V^2$, $s\gg t$ behavior of the form
\beq
T_{ab\to cd}(s,t) =  g^2_V(\delta_{ab}\delta_{cd}-\delta_{ad}\delta_{bc})\, \frac{s}{t-M_V^2}\, + \dots ,
\eeq
which gives
\beq
\calA(s)=(2w_0+w_1-w_2)\frac{g^2_V s}{M_V^2},\qquad s\gg M_V^2\,.
\eeq
Notice that, unlike for the pomeron exchange, this asymptotic behavior is purely real and thus does not cause a divergence in the integrals over the cuts. Nevertheless, there is a nonvanishing contribution from the big circle, so that the RHS of our sum rule gets an extra contribution:
\beq
c_\infty= g^2_V v^2/M_V^2\,.
\eeq
Since this contribution is positive, it does not modify our arguments given above that $a>1$ needs an increase in the $I=2$ cross section.

Let us now list a few physical scenarios for which our sum rule will be of interest. What are the possibilities for light scalars with $a\ne1$? One example is to have a light dilaton \cite{Goldberger:2007zk}. In this case $a=v/f$, where $f$ is the scale of spontaneous conformal symmetry breaking of which the dilaton is the Goldstone boson. Perturbative intuition may suggest that $v < f$ (and hence $a<1$) since all vevs contribute to $f$ but not necessarily to  $v$. However, the fact that our some rule is not sign definite indicates that this reasoning may be too naive. 

Spontaneous breaking of conformal symmetry producing a light dilaton does not look easy to come by without supersymmetry (although see \cite{Rattazzi-talk}). A more generic possibility is to have a light Higgs as a pseudo-Goldstone boson of spontaneously broken \emph{global} symmetry at a scale $f$ somewhat higher than $v$. 
Generic features of this scenario have been distilled in the SILH framework \cite{Giudice:2007fh}. To first order in $\xi=v^2/f^2$, the coupling $a$ in SILH is related by
\beq
a=1-c_H \xi/2
\eeq
to the coefficient $c_H$ of a particular dimension 6 effective operator written in terms of a composite Higgs doublet field,
\beq
\mathcal O_H=\frac{c_H}{2f^2}\del^\mu(H^\dagger H)\del_\mu(H^\dagger H)\,,
\label{e.OH}
\eeq
which renormalizes the Higgs boson wave function. Basic group theory implies \cite{Low:2009di} that for all `normal' cosets $c_H>0$, so that this effect always suppresses the coupling: $a<1$. By `normal' here we mean cosets $\mathcal G/\mathcal H$ based on a \emph{compact} global symmetry group $\mathcal G$. Only such cosets are allowed if the UV completion is described by a quantum field theory, since at high energies, where the full symmetry $\mathcal G$ is expected to be restored, the symmetry currents of a non-compact global symmetry will have non sign-definite two point functions, which is inconsistent with unitarity. However, non-compact coset can be perfectly healthy as a low energy effective theory (as long as the unbroken group $\mathcal H$ is compact). In fact, string theory low-energy effective actions (supergravities) often contain scalar fields living in non-compact cosets \cite{Rattazzi1}. And if a pseudo-Goldstone Higgs boson lives in a non-compact coset, like SO(4,1)/SO(4), then the wave function effect sign is reversed, and one obtains $a>1$ \cite{Rattazzi1,Low}. 

Our sum rule should apply both to compact and non-compact cosets. For $a<1$ the composite Higgs boson `under-unitarizes' the longitudinal WW scattering, as the scattering amplitude (\ref{e.lowAmp}) continues to grow at $s\gg m_h^2$, although with a reduced coefficient. Our sum rule then suggests that unitarization is  completed by heavier resonances of isospin 0 and 1. In fact, unitarization by isospin 1 heavy vectors is a possibility much discussed in the literature, although isospin 0 heavy scalars is also an option. On the other hand, for non-compact cosets $a>1$ and one can say that the Higgs boson `over-unitarizes'. In this case the sum rules suggests isospin 2 resonances which could restore the unitarization balance.

We can rephrase our sum rule in terms of the total $\pi \pi$ scattering cross section for various charge combinations,    
\beq
\hspace{-0.3cm}
1 - a^2  = {v^2 \over \pi}  \int_0^\infty {ds \over s}  
\left (\sigma_{00}^{\rm tot} +  \sigma_{+0}^{\rm tot}  -  2 \sigma_{++}^{\rm tot} \right )
= {v^2 \over \pi}  \int_0^\infty {ds \over s}  
\left (2 \sigma_{+-}^{\rm tot} -  \sigma_{00}^{\rm tot}  -  \sigma_{+0}^{\rm tot} \right )
=  {v^2 \over \pi}  \int_0^\infty {ds \over s}     \left (  \sigma_{+-}^{\rm tot}  - \sigma_{++}^{\rm tot} \right ). 
\label{eq:reduced}
\eeq 
The last form is related to the sum rule 
\beq
c_H=c_\infty+ {f^2 \over \pi}  \int {ds \over s}     \left (  \sigma_{+-}^{\rm tot}  - \sigma_{++}^{\rm tot} \right )
\eeq
derived in Ref.~\cite{Low:2009di} for the coefficient of the SILH effective operator (\ref{e.OH}), by very similar methods based on analyticity and crossing. In particular, Ref.~\cite{Low:2009di} identified correctly that $c_\infty>0$ may arise for perturbative UV-completions containing t-channel vector exchanges. Our sum rule is stronger as it uses the full power of the isospin and shows that the contact term $c_\infty$ is likely unnecessary for strongly coupled UV completions.
 
 {\bf UV completion of $a>1$ via a Scalar Quintuplet.}   
 Let us now examine the situation $a>1$ more closely but without necessarily committing ourselves to the motivated scenarios (dilaton, Higgs as a PGB) discussed above. Our sum rule demonstrates that $a > 1$ is possible only when there is an enhancement in the $I = 2$ channel of longitudinal electroweak gauge boson scattering, and the most natural way to produce such an enhancement is via an s-channel isospin 2 resonance. The sum rule fixes the coupling of such a resonance to the $W$ and $Z$ bosons. As we will now explicitly demonstrate, the value of this coupling can be understood by using the familiar concept of unitarization.

Consider an isospin-2 scalar quintuplet  $Q = (Q^{++},Q^+,Q^0,Q^-,Q^{--})$ with mass $m_Q$ coupled in the isospin invariant way (in the limit $g' \to 0$) to the electroweak gauge bosons:
\beq
\label{e.qvv}
\cl_{Q} =  {g_{Q} \over v}  \left \{  \sqrt{2 \over 3} Q^0  \left (  m_W^2 W_\mu^+ W_\mu^-  -  m_Z^2 Z_\mu^2 \right )  
+  \left ( Q^{++} m_W^2 W_\mu^- W_\mu^-  
+  \sqrt{2}  Q^{+} m_W m_Z  W_\mu^- Z_\mu   + \hc \right ) \right \}.
\eeq 
In the presence of the quintuplet the basic pion amplitude takes the form  \cite{Alboteanu:2008my}
\beq
A(s,t,u) =   {s \over v^2}\left(1 - a^2 {s \over s - m_h^2}\right)   + {g_{Q}^2 \over v^2} \left ( 
 {s^2 \over 3 (s - m_Q^2)}  -  {t^2\over 2 (t - m_Q^2)} -  {u^2\over 2 (u - m_Q^2)}
  \right ). 
\eeq 
In the considered case $a>1$, at the energies  $s\gg m_h^2$ the scattering amplitude still grows with $s$ but with a negative coefficient (`over-unitarization'). However, if the resonance coupling is fixed at 
\beq  \label{e.gq} g_{Q}^2 = {6 \over 5} \left (a^2-1\right)\,,
 \eeq
then $A(s,t,u)$ stops growing at $s\gg M_Q^2$ and the perturbative unitarity is restored. The cross section corresponding to this coupling is precisely the one predicted by the sum rule. 
 
This raises a curious question: is it possible to embed the above effective theory into a complete renormalizable model valid up to arbitrary high scales and having $a>1$?
The answer is yes. 
The Higgs sector of the minimal model with this property is a $3 \times 3$ matrix of scalar fields  $\Phi$ transforming as $(3,3)$ under global $SU(2) \times SU(2)$.\footnote{For a general discussion of extended Higgs sectors with custodial symmetry, see \cite{Low:2010jp}.} Under $SU(2)_V$,  $\Phi$ decomposes as $\mathbf{1}+\mathbf{3}+\mathbf{5}$, where $\mathbf{3}$ will be the eaten Goldstones $\pi^a$, while $\mathbf{1}$ and $\mathbf{5}$ are the $h$ and $Q$ of the above example. As usual, the electroweak symmetry is the subgroup of the global symmetry, with $SU(2)_L$ identified with one $SU(2)$ factor, and $U(1)_Y$ realized as the $T^3$ generator of the other $SU(2)$.  The electroweak symmetry is broken to $U(1)_{\rm em}$ by the vev $\la \Phi \ra = {v \over 2 \sqrt 2} I_{3 \times 3}$, which also breaks the global symmetry to the diagonal isospin $SU(2)_V$.  
The singlet $h$, quintuplet $Q$, and the triplet of Goldstones $\pi^a$ (eaten by W and Z) are embedded into $\Phi$ as\footnote{
$\Phi$ is defined to be a real $3\times 3$ matrix in the adjoint (real) basis. In the charge basis which we're using here the reality condition  translates to $\Phi^\dagger = (RR^T) \Phi (R R^T)$ where $R$ is defined in  \eref{AdjointToCharge}. This in turn corresponds  to $\Phi$ being equal to its complex conjugate under reflection with respect to the central $\Phi_{22}$ element.} 
\beq
\Phi =  \left ( \ba{ccc}
{v \over 2 \sqrt{2}}   + {1 \over  \sqrt 3} h - {1 \over \sqrt 6} Q^0+ {i \over \sqrt 2}  \pi^0 &  - {1 \over \sqrt{2}} (Q^+ + i \pi^+)     & - Q^{++} \\ 
 - {1 \over \sqrt{2}} (Q^- + i \pi^-)  & {v \over 2 \sqrt{2}}   + {1 \over  \sqrt 3} h +  {\sqrt{2\over 3}} Q^0 & - {1 \over \sqrt{2}} (Q^+ - i \pi^+)  \\
- Q^{--} & - {1 \over \sqrt{2}} ( Q^- -  i \pi^-) &   {v \over 2 \sqrt{2}}   + {1 \over  \sqrt 3} h - {1 \over \sqrt 6} Q^0 - {i \over \sqrt 2}  \pi^0 
\ea \right ).
\eeq  
The Higgs sector  sector Lagrangian is
\beq
\label{e.HiggsSector}
\cl_{\rm Higgs} = {1 \over 2} \tr \left [D_\mu \Phi^\dagger D_\mu \Phi \right ]  - V(\Phi), 
\quad
V(\Phi) =  {1 \over 2} m^2  \tr [\Phi^\dagger \Phi] +  \lambda_1  \tr [\Phi^\dagger \Phi \Phi^\dagger \Phi] +  \lambda_2  \tr [\Phi^\dagger \Phi]  \tr [\Phi^\dagger \Phi] 
\eeq 
where $D_\mu \Phi = \pa_\mu \Phi - i g L_\mu^a T^a \Phi + i g' B_\mu  \Phi T^3$ and the generators $T^a$ are defined in \eref{su2gen_charge}.  
The singlet and quintuplet masses that follow are given by 
\beq
m_h^2 =   (\lambda_1 + 3 \lambda_2) v^2, 
\qquad 
m_Q^2 = \lambda_1 v^2,  
\eeq  
and can be dialed independently. It is straightforward to demonstrate that  \eref{HiggsSector} leads to the singlet and quintuplet interactions with electroweak gauge bosons of the form \eref{hvv} and \eref{qvv} with the couplings $a =\sqrt{8/3}$, $g_{Q} = \sqrt{2}$.   
Clearly, $a > 1$ and the relation \eref{gq} is fulfilled.  

Note that while the minimal custodially invariant renormalizable model with $a>1$ predicts $a =\sqrt{8/3}$, in more general renormalizable models this relation can be easily relaxed. The coupling of the isospin singlet to $W$ and $Z$ can be reduced if $h$ mixes with a gauge singlet $N$, for example via a term $N \tr [\Phi^\dagger \Phi]$ in the potential. In that case two isospin singlet mass eigenstates with $a < \sqrt{8/3}$ are present in the spectrum, however one may reside at the TeV scale and not be easily discoverable at the LHC. The same goes for the model where $\Phi$ is accompanied by the usual Higgs double $H$, in which case 2 isospin singlets, 1 triplet, and 1 quintuplet are present in the physical spectrum. This last case is known in the literature as the Georgi-Machacek model \cite{Georgi:1985nv}.

{\bf Comments on Higgs Phenomenology.}  
The increased  coupling to the electroweak gauge bosons may have several effects on the Higgs phenomenology at  the LHC, where by Higgs we mean the isospin singlet $h$  (see also \cite{Low:2010jp}). 
First of all, the Higgs decay width into WW and ZZ is enhanced: 
\beq
{\Gamma(h \to WW) \over \Gamma_{SM}(h \to WW) } =  {\Gamma(h \to ZZ) \over \Gamma_{SM}(h \to ZZ) } = a^2 . 
\eeq    
By the same token, the vector-boson-fusion Higgs production mode is enhanced compared to the SM. 

The Higgs decay width into photons can be modified by two separate effects.   
First, the contribution of the W boson loop to the $h \to \gamma \gamma$  amplitude, which is the dominant one in the SM, gets enhanced by $a$. 
However, if the model is UV completed with a scalar quintuplet, the symmetries allow  the latter to couple to the Higgs as 
\beq
\cl_{hQQ} = 
- 2 g_{h Q Q}  m_Q^2  {h \over v} \left (|Q^{++}|^2 + |Q^{+}|^2 + {1 \over 2} (Q^0)^2 \right )  .  
\eeq 
In particular, in the minimal renormalizable model \eref{HiggsSector} one finds $g_{hQQ} =  \sqrt{2 \over 3}{ m_h^2 + 2 m_Q^2 \over m_Q^2}$.
Thus, the charged members of the quintuplet can enter  the $h \to \gamma \gamma$  loop amplitude. 
Assuming only these two factors affect the decay (in particular, assuming that $h$ couples  to fermions as in the SM), for $m_h \sim 125 $ GeV the width is modified approximately as 
 \beq
 {\Gamma(h \to \gamma \gamma) \over \Gamma_{SM}(h \to \gamma \gamma) } \approx  \left ({a -  {2 \over 9} -  {5 \over 24}  g_{h Q Q}  \over 7/9 }   \right )^2. 
\eeq 
Since the result depends on $g_{h Q Q}$, which is a free parameter not constrained by unitarity arguments,  we cannot make a definite prediction about the $h\to \gamma \gamma $ width. One should also remember that the Higgs production rates in all channels  strongly depend on the  Higgs couplings to the SM fermions, which may also be modified in the present context.

{\bf Summary.}
We argued that an observation of enhanced coupling of  the Higgs boson to the W and Z bosons implies the enhancement of the longitudinal gauge boson scattering cross section in the isospin-2 channel, such as via a quintuplet of narrow resonances (including doubly charged ones) coupled to WW, WZ, and ZZ. 
 

\section*{Acknowledgements}
We thank Roberto Contino for important discussions and collaboration at the early stage of this project.  
We are grateful to Yuri Dokshitzer and Riccardo Rattazzi for useful discussions. We are greatful to Marc Knecht for pointing out 
\cite{Olsson}. This work is supported in part by the European Program ÒUnification in the LHC EraÓ, contract PITN-GA-2009-237920 (UNILHC). We acknowledge funding from the \'Emergence-UPMC-2011 research program. 


\appendix 
\renewcommand{\theequation}{\Alph{section}.\arabic{equation}} 
\setcounter{section}{0} 
\setcounter{equation}{0}

\section{Appendix: Isospin amplitudes}  

Everywhere in this paper we assumed that the electroweak symmetry breaking sector respects an $SU(2)\times SU(2)$ global symmetry, spontaneously broken to the global $SU(2)_V$ referred to as {\em custodial isospin}. As is well known, custodial isospin ensures the phenomenologically successful relation $m_W/m_Z \approx \cos \theta_W$, and is therefore expected to remain a good approximate symmetry in any realistic extension of the SM. 
The three ``pions'' $\pi^a$ (the Goldstone bosons that become longitudinal components of  W and Z) transform as the triplet representation of custodial isospin.   
By isospin and crossing symmetry, all pion 2-to-2 scattering amplitude can be expressed by one function of the kinematic variables $A(s,t,u)$ in the form given in \eref{basic}. 

Before we give the isospin decomposition of the pion amplitude, we first summarize our  conventions for $SU(2)$ generators.  
The triplet representation in the real (adjoint)  basis $\pi^a$ can be defined using the structure constants: $(T^b)_{ac} = i  \eps^{a b c}$. 
It is often more convenient to work in the charge eigenstate basis where $T^3$ is diagonal,   
\beq
\label{e.AdjointToCharge}
\bvec \pi^+  \\ \pi^0  \\ \pi^- \evec  =  R \cdot \bvec \pi^1 \\ \pi^2 \\ \pi^3 \evec ,
\qquad 
R = \left ( \ba{ccc} 
1/\sqrt 2& - i/\sqrt{2}&0  
\\
0& 0& 1
\\ 
1/\sqrt 2& i/\sqrt{2}& 0 
\ea  \right).  
\eeq 
The generators transform as $T^a \to R T^a R^\dagger$, hence in the charge basis\footnote{Our generators $T^{1,2}$ differ by a sign convention from those used in most quantum mechanics textbooks. The reason is that, to arrive at the standard form, one needs to define the charge basis such that $\pi^+ = - (\pi^-)^\dagger$ or $\pi^0 = -(\pi^0)^\dagger$, which is rather awkward in the field theory context. Consequently, our Clebsch-Gordan coefficient in \eref{clebsch} also differ by a sign convention from the standard ones. }  
\beq\label{e.su2gen_charge}
T^1 =  {1 \over \sqrt 2}\left ( \ba{ccc}  0 & -1 & 0 \\ -1 & 0 & 1\\ 0 & 1 & 0 \ea \right ), 
\quad 
T^2 =   {1 \over \sqrt 2} \left ( \ba{ccc}  0 & i & 0 \\ -i & 0 & -i\\ 0 & i & 0 \ea \right ), 
\quad 
T^3 =   \left ( \ba{ccc}  1 & 0 & 0 \\ 0 & 0 & 0 \\ 0 & 0 & -1 \ea \right ).  
\eeq 
Using \eref{AdjointToCharge}, the pion amplitudes in the charge basis can be easily derived  from \eref{basic}, 
\begin{align}
\label{e.TtoA}
&T_{\pi^0\pi^0 \to \pi^+ \pi^-} =    A_s,
&&T_{\pi^0\pi^0 \to \pi^0 \pi^0} =   A_s + A_t  + A_u 
\nn
&T_{\pi^\pm\pi^0 \to \pi^\pm \pi^0} =   A_t\,, 
&& T_{\pi^+\pi^- \to \pi^+ \pi^-}  =  A_s + A_t \,,
\nn 
& T_{\pi^\pm\pi^0 \to \pi^0 \pi^\pm}  =  A_u, 
&&T_{\pi^\pm \pi^\pm \to \pi^\pm \pi^\pm}  =   A_t + A_u 
\end{align}
We will also need the amplitudes in the isospin basis, which diagonalize the S-matrix:
\beq
\ostate{I,m} S \qstate{I',m'} = T_I(s,t) \delta_{I I'}  \delta_{m m'} 
\eeq 
The Wigner-Eckart theorem says that these amplitudes depend only $I$ and not on $m$. 
Isospin decomposition of the ${\bf 3} \times {\bf 3}$ product representation is given by 
\beq
\nonumber 
\hspace{-0.5cm}
\bvec \qstate{2,2}\\ \qstate{2,1} \\ \qstate{2,0} \\ \qstate{2,-1}\\ \qstate{2,-2} \evec 
= \bvec  \qstate{\pi^+\pi^+} \\ {1 \over \sqrt 2} \left (\qstate{\pi^+\pi^0}  +   \qstate{\pi^0\pi^+} \right ) \\  {1 \over \sqrt 6} \left (  \qstate{\pi^+\pi^- } + \qstate{\pi^-\pi^+ }  -  2 \qstate{\pi^0\pi^0} \right )   \\  
{1\over \sqrt 2}\left (\qstate{\pi^-\pi^0}  +   \qstate{\pi^0\pi^-} \right )   \\   \qstate{\pi^-\pi^-} \evec,  
\quad 
\bvec 
\qstate{1,1} \\ \qstate{1,0} \\ \qstate{1,-1}
\evec  = 
\bvec  {1 \over \sqrt 2} \left (\qstate{\pi^+\pi^0}  -   \qstate{\pi^0\pi^+} \right )   \\  {1 \over \sqrt 2} \left (\qstate{\pi^+\pi^-} -  \qstate{\pi^-\pi^+} \right )  \\  {1\over \sqrt 2}\left (\qstate{\pi^-\pi^0}  -   \qstate{\pi^0\pi^-} \right )   \evec,  
 \eeq 
 \vspace{-0.3cm}
\beq
\label{e.clebsch}
 \qstate{0,0} = {1 \over \sqrt 3} \left ( \qstate{\pi^+\pi^-} +  \qstate{\pi^-\pi^+} + \qstate{\pi^0\pi^0} \right ). 
\eeq 
Eq.\ (\ref{e.clebsch}) together with  \eref{TtoA} immediately lead to the isospin amplitudes given in \eref{isoAmp}.



\begin{thebibliography}{99}


\bibitem{Low:2009di}
I.~Low, R.~Rattazzi and A.~Vichi,
``Theoretical Constraints on the Higgs Effective Couplings,''
JHEP {\bf 1004} (2010) 126
\hhref{0907.5413}.


\bibitem{Giudice:2007fh}
G.~F.~Giudice, C.~Grojean, A.~Pomarol and R.~Rattazzi,
``The Strongly-Interacting Light Higgs,''
JHEP {\bf 0706} (2007) 045
\hhref{hep-ph/0703164}.

\bibitem{cmsgaga}  
S.~Chatrchyan {\it et al.} [CMS Collaboration],
``Search for the Standard Model Higgs Boson Decaying into Two Photons in PP Collisions at $\sqrt{s} =7$ TeV,'' \hhref{1202.1487}.


\bibitem{Adams:2006sv}
A.~Adams, N.~Arkani-Hamed, S.~Dubovsky, A.~Nicolis and R.~Rattazzi,
``Causality, Analyticity and an IR Obstruction to UV Completion,''
JHEP {\bf 0610} (2006) 014
\hhref{hep-th/0602178}.

\bibitem{Olsson}
M.~G.~Olsson,
  ``Low-Energy $p$-Wave $\pi$-$\pi$ Interaction,''
  Phys.\ Rev.\ {\bf 162}, 1338 (1967).

\bibitem{Adler}
  S.~L.~Adler,
  ``Sum rules for the axial vector coupling constant renormalization in $\beta$ decay,''
  Phys.\ Rev.\  {\bf 140}, B736 (1965)
  [Erratum-ibid.\  {\bf 149}, 1294 (1966)]
  [Erratum-ibid.\  {\bf 175}, 2224 (1968)].
  
\bibitem{Nieves:2011gb}
J.~Nieves, A.~Pich and E.~Ruiz Arriola,
``Large-Nc Properties of the $\rho$ and $f_{0}$(600) Mesons from Unitary Resonance Chiral Dynamics,''
Phys.\ Rev.\ D {\bf 84} (2011) 096002 \hhref{1107.3247}

\bibitem{Pomeron}
J.~R.~Forshaw and D.~A.~Ross,
``Quantum Chromodynamics and the Pomeron,''
Cambridge Univ. Press, 1997.

\bibitem{Goldberger:2007zk}
W.~D.~Goldberger, B.~Grinstein and W.~Skiba,
``Light Scalar at LHC: the Higgs Or the Dilaton?,''
Phys.\ Rev.\ Lett.\ {\bf 100} (2008) 111802 \hhref{0708.1463}.

\bibitem{Rattazzi-talk} R.~Rattazzi ``The naturally light dilaton or How to break dilations
spontaneously and naturally", talk at Planck 2010, CERN \href{http://indico.cern.ch/getFile.py/access?contribId=163&resId=0&materialId=slides&confId=75810}{[slides]}.

\bibitem{Rattazzi1} R.~Rattazzi ``EWSB after the first hints of a Higgs", Workshop "Higgs search confronts theory", Zurich, 9-11 Jan 2012 \href{http://www.zpw.ethz.ch/2012/slides/rattazzi_zpw2012.pdf}{[slides]}.

\bibitem{Low} 
I.~Low ``A minimally symmetric Higgs boson", Workshop on Strongly Coupled Physics Beyond the Standard Model, ICTP, Trieste, 25-27 Jan 2012 \href{http://cdsagenda5.ictp.trieste.it/askArchive.php?base=agenda&categ=a11229&id=a11229s1t14/slides}{[slides]}.


\bibitem{Alboteanu:2008my}
A.~Alboteanu, W.~Kilian and J.~Reuter,
``Resonances and Unitarity in Weak Boson Scattering at the LHC,''
JHEP {\bf 0811} (2008) 010
\hhref{0806.4145}. 
R.~Contino, D.~Marzocca, D.~Pappadopulo and R.~Rattazzi,
``On the Effect of Resonances in Composite Higgs Phenomenology,''
JHEP {\bf 1110} (2011) 081 \hhref{1109.1570}.



\bibitem{Low:2010jp}
I.~Low and J.~Lykken,
``Revealing the Electroweak Properties of a New Scalar Resonance,''
JHEP {\bf 1010} (2010) 053 \hhref{arXiv:1005.0872}.



\bibitem{Georgi:1985nv} 
  H.~Georgi and M.~Machacek,
  ``Doubly Charged Higgs Bosons,''
  Nucl.\ Phys.\ B {\bf 262}, 463 (1985).
  J.~F.~Gunion, R.~Vega and J.~Wudka,
  ``Higgs triplets in the standard model,''
  Phys.\ Rev.\ D {\bf 42}, 1673 (1990).






\end{thebibliography}
\end{document}